\documentclass[%
 reprint,showpacs,
 superscriptaddress,
 amsmath,amssymb,
 aps,
 prl,
]{revtex4-1}

\usepackage{graphicx}
\usepackage{dcolumn}
\usepackage{bm}
\usepackage[
   bookmarks=true,
   colorlinks=true,
   hyperindex=true,
   citecolor=blue,
   linkcolor=blue,
   urlcolor=blue
]{hyperref}

\begin{document}

\title{Test of Equivalence Principle at $10^{-8}$ Level by a Dual-species Double-diffraction Raman Atom Interferometer}

\author{Lin Zhou}
\affiliation{State Key Laboratory of Magnetic Resonance and Atomic and Molecular Physics, Wuhan Institute of Physics and Mathematics, Chinese Academy of Sciences, Wuhan 430071, China}
\affiliation{Center for Cold Atom Physics, Chinese Academy of Sciences, Wuhan 430071, China}
\author{Shitong Long}
\affiliation{State Key Laboratory of Magnetic Resonance and Atomic and Molecular Physics, Wuhan Institute of Physics and Mathematics, Chinese Academy of Sciences, Wuhan 430071, China}
\affiliation{Center for Cold Atom Physics, Chinese Academy of Sciences, Wuhan 430071, China}
\affiliation{University of Chinese Academy of Sciences, Beijing 100049, China}
\author{Biao Tang}
\affiliation{State Key Laboratory of Magnetic Resonance and Atomic and Molecular Physics, Wuhan Institute of Physics and Mathematics, Chinese Academy of Sciences, Wuhan 430071, China}
\affiliation{Center for Cold Atom Physics, Chinese Academy of Sciences, Wuhan 430071, China}
\author{Xi Chen}
\affiliation{State Key Laboratory of Magnetic Resonance and Atomic and Molecular Physics, Wuhan Institute of Physics and Mathematics, Chinese Academy of Sciences, Wuhan 430071, China}
\affiliation{Center for Cold Atom Physics, Chinese Academy of Sciences, Wuhan 430071, China}
\author{Fen Gao}
\affiliation{State Key Laboratory of Magnetic Resonance and Atomic and Molecular Physics, Wuhan Institute of Physics and Mathematics, Chinese Academy of Sciences, Wuhan 430071, China}
\affiliation{Center for Cold Atom Physics, Chinese Academy of Sciences, Wuhan 430071, China}
\author{Wencui Peng}
\affiliation{State Key Laboratory of Magnetic Resonance and Atomic and Molecular Physics, Wuhan Institute of Physics and Mathematics, Chinese Academy of Sciences, Wuhan 430071, China}
\affiliation{Center for Cold Atom Physics, Chinese Academy of Sciences, Wuhan 430071, China}
\author{Weitao Duan}
\affiliation{State Key Laboratory of Magnetic Resonance and Atomic and Molecular Physics, Wuhan Institute of Physics and Mathematics, Chinese Academy of Sciences, Wuhan 430071, China}
\affiliation{Center for Cold Atom Physics, Chinese Academy of Sciences, Wuhan 430071, China}
\affiliation{University of Chinese Academy of Sciences, Beijing 100049, China}
\author{Jiaqi Zhong}
\affiliation{State Key Laboratory of Magnetic Resonance and Atomic and Molecular Physics, Wuhan Institute of Physics and Mathematics, Chinese Academy of Sciences, Wuhan 430071, China}
\affiliation{Center for Cold Atom Physics, Chinese Academy of Sciences, Wuhan 430071, China}
\author{Zongyuan Xiong}
\affiliation{State Key Laboratory of Magnetic Resonance and Atomic and Molecular Physics, Wuhan Institute of Physics and Mathematics, Chinese Academy of Sciences, Wuhan 430071, China}
\affiliation{Center for Cold Atom Physics, Chinese Academy of Sciences, Wuhan 430071, China}
\author{Jin Wang}
\email{wangjin@wipm.ac.cn}
\affiliation{State Key Laboratory of Magnetic Resonance and Atomic and Molecular Physics, Wuhan Institute of Physics and Mathematics, Chinese Academy of Sciences, Wuhan 430071, China}
\affiliation{Center for Cold Atom Physics, Chinese Academy of Sciences, Wuhan 430071, China}
\author{Yuanzhong Zhang}
\affiliation{Institute of Theoretical Physics, Chinese Academy of Sciences, Beijing 100190, China}
\author{Mingsheng Zhan}
\email{mszhan@wipm.ac.cn}
\affiliation{State Key Laboratory of Magnetic Resonance and Atomic and Molecular Physics, Wuhan Institute of Physics and Mathematics, Chinese Academy of Sciences, Wuhan 430071, China}
\affiliation{Center for Cold Atom Physics, Chinese Academy of Sciences, Wuhan 430071, China}

\date{\today}

\begin{abstract}
We report an improved test of the weak equivalence principle by using a simultaneous $^{85}$Rb-$^{87}$Rb dual-species atom interferometer. We propose and implement a four-wave double-diffraction Raman transition scheme for the interferometer, and demonstrate its ability in suppressing common-mode phase noise of Raman lasers after their frequencies and intensity ratios are optimized. The statistical uncertainty of the experimental data for E\"{o}tv\"{o}s parameter $\eta$ is $0.8\times10^{-8}$ at 3200 s. With various systematic errors corrected the final value is $\eta=(2.8\pm3.0)\times10^{-8}$. The major uncertainty is attributed to the Coriolis effect.
\end{abstract}



\maketitle

The equivalence principle including the weak equivalence principle (WEP), also known as universality of free fall, is one of the two assumptions of Einstein's general relativity. Theories which try to unify gravity and the standard model, generally require violation of WEP \cite{unifytheory}. To explore the applicable extent of  WEP and to help the birth of new quantum gravity theories, it is very important to precisely test WEP both with macro objects and with microscopic particles. WEP has been tested experimentally with large objects by lunar laser ranging \cite{JGW2004} and torsion balances \cite{SS2008} at $10^{-13}$ level, while with atoms it is tested only at $10^{-7}$ level.

The test with atoms relies on atom interferometry which has been developed for over 20 years \cite{RMP2009} and has been widely used in measurements of gravity \cite{YB2013} and its gradient \cite{FS2014}, the Newtonian gravitational constant \cite{GR2014}, gravitational redshift\cite{HM2010}, and post-Newtonian gravity\cite{HM2008}. Fray \emph{et al}. \cite{SF2004} performed the first atom based WEP test using an atom interferometer (AI) with an E\"{o}tv\"{o}s value of $\eta = (1.2\pm1.7)\times10^{-7}$  by measuring the gravitational accelerations of the isotopic $^{85}$Rb and $^{87}$Rb atoms. Ten years later Bonnin \emph{et al}. \cite{AB2013} reported the same test to a similar accuracy of $\eta = (1.2\pm3.2)\times10^{-7}$ by using simultaneous dual-species ($^{85}$Rb and $^{87}$Rb) AIs. A non-isotopic pair of atoms $^{87}$Rb and $^{39}$K was also used recently by Schlippert \emph{et al}. \cite{DS2014}, they tested WEP with $\eta = (0.3\pm5.4)\times10^{-7}$. In addition, the bosonic and fermionic isotopes of strontium atoms were also used to test WEP, the value is $(0.2\pm1.6)\times10^{-7}$ \cite{MG2014}.

On the other hand, current single-species AI technique has reached very high resolution \cite{SD2007,SMD2013}, which could principally push the AI based WEP test to a much higher accuracy than $10^{-7}$. The main obstacles are complex noise that is difficult to be common-mode rejected, and crosstalk of different laser frequencies in a dual-species AI.

Here we propose a simultaneous dual-species double-diffraction Raman AI and demonstrate a new WEP test with it. We design and realize a four-wave double-diffraction Raman transition (4WDR) scheme by carefully selecting the frequencies and intensity ratio of Raman beams to avoid the crosstalk among different lasers. The 4WDR is based on the single-species double-diffraction Raman AI \cite{TL2009,NM2010}, but extended to two species ($^{85}$Rb and $^{87}$Rb).

\begin{figure}[htbp]
\centering
\includegraphics{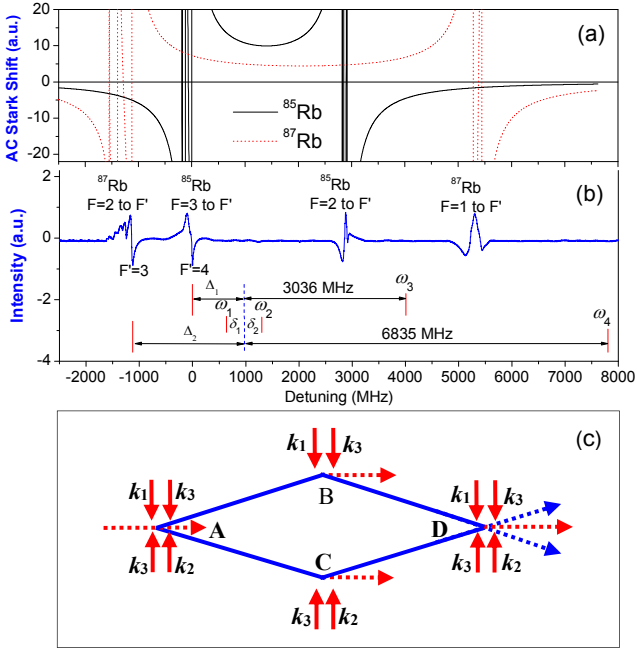}
\caption{(color online) Schematic diagram of the 4WDR scheme. (a) AC Stark shift spectrum of rubidium atoms. (b) Lasers with frequencies of $\omega_{i} (i=1\sim4)$ are used as Raman beams for $^{85}$Rb - $^{87}$Rb dual-species AI; $\delta_{1}$ is the detuning of $\omega_{1}$ , $\delta_{2}$ is the detuning of $\omega_{2}$ . $\omega_{1}$ and $\omega_{2}$ are detuned to blue side of transitions $F=3$ to $F^{\prime}=4$ of $^{85}$Rb and $F=2$ to $F^{\prime}=3$ of $^{87}$Rb. (c) Diagram of a double-diffraction Raman AI using $k_{1}, k_{2}$, and $k_{3}$, where the blue lines are the paths of atoms in ground state $F=2$, while the red dashed lines are for excited state $F=3$. }
\label{fig:fig1}
\end{figure}

Our 4WDR scheme is illustrated in Fig.\ref{fig:fig1}. Raman beams for the dual-species AI are composed of four lasers with frequencies of $\omega_{i}$ and wave vectors $\emph{k}_{i}(i=1\sim4)$. $\omega_{1}$ and $\omega_{2}$  are used as shared Raman beams for $^{85}$Rb and $^{87}$Rb atoms, while $\omega_{3}$ and $\omega_{4}$ are for $^{85}$Rb and $^{87}$Rb respectively. Pairs $(\omega_{1}, \omega_{3})$ and $(\omega_{2}, \omega_{3})$ are for double-diffraction Raman transition of $^{85}$Rb AI, while $(\omega_{1}, \omega_{4})$ and $(\omega_{2}, \omega_{4})$ are for $^{87}$Rb.

In the 4WDR scheme the frequencies and intensity ratios of Raman beams are chosen to meet the following requirements: 1) the four frequencies are far off-resonant to all of the resonance lines of rubidium isotopes; 2) the intensities of two chirp lasers ($\omega_{1}$ and $\omega_{2}$) are equal, to ensure that the corresponding Rabi frequencies of two counter-propagated wave vectors in each double-diffraction Raman transition are equal, and atoms recoil to two interference paths with the same probability; 3) the corresponding Rabi frequencies of different species AIs are the same; 4) for dual-species Raman transitions, the total AC Stark shift caused by four Raman beams are zero.

To find the optimal parameters, we calculate the AC Stark shift spectrum of rubidium atoms (see  Fig.\ref{fig:fig1}(a)). To cancel AC Stark shifts in both species AIs, some Raman frequencies should lie between the cooling laser ($F=3$ to $F^{\prime}$) and repumping laser ($F=2$ to $F^{\prime}$) for $^{85}$Rb atoms. The frequencies of Raman beams we selected are shown in Fig.\ref{fig:fig1}(b), they satisfy the following relation
\begin{equation*}
\omega_{1}+\delta_{1}=\omega_{2}-\delta_{2}=\omega_{3}-3.036 GHz=\omega_{4}-6.835 GHz
\end{equation*}
where $\delta_{i}$ is the detuning of $\omega_{i} (i =1, 2)$. $\delta_{1}$=-$\delta_{2}$=$\upsilon\emph{k}_{eff}/2\pi$, where $\upsilon$ is the projection of atomic velocity along the direction of wave vector, $\emph{k}_{eff} = \emph{k}_{1} + \emph{k}_{2} + 2\emph{k}_{3}$ (for $^{85}$Rb) or $\emph{k}_{1} + \emph{k}_{2} + 2\emph{k}_{4}$ (for $^{87}$Rb) are the effective wave vectors of the Raman lasers, $\upsilon\emph{k}_{eff}/2\pi$ equals Doppler shift of atoms. $\omega_{1}$ and $\omega_{2}$ are detuned $\Delta_{1}$=971 MHz and $\Delta_{2}$=2097 MHz respectively to the blue sides of transitions $F=3$ to $F^{\prime}=4$ of $^{85}$Rb and $F=2$ to $F^{\prime}=3$ of $^{87}$Rb. Shown in the up row of Fig.\ref{fig:fig1}(b) is a polarization spectrum of rubidium atoms \cite{Peng2014} for reference. By fixing the above frequency locations we then decide the intensities. We find that the optimal intensity ratios of four Raman beams are $I_{1}:I_{2}:I_{3}:I_{4}=1.0:1.0:3.1:14.3$, where $I_{i}$ is the intensity of $\omega_{i} (i=1\sim4)$.

A pronounced advantage of the 4WDR scheme is its capability to suppress the common mode phase noise of Raman lasers. This can be seen by writing the total phase shift \cite{MK1992} of a single-species (taking $^{85}$Rb as an example and shown in Fig.\ref{fig:fig1}(c)) double-diffraction Raman AI
\begin{equation*}
\Delta\varphi = k_{eff}gT^{2}+\Delta\varphi_{B}+\Delta\varphi_{C}-\Delta\varphi_{A}-\Delta\varphi_{D}
\end{equation*}
where \emph{T} is the time interval of $\pi/\sqrt{2} - \sqrt{2}\pi - \pi/\sqrt{2}$ Raman pulse sequence \cite{TL2009}, $\Delta\varphi_{j}(\emph{j}=A \sim D)$ is the initial phase shift at site \emph{j}. Since Raman pairs ($k_{1}, k_{3}$) and ($k_{2}, k_{3}$) supply recoil momentum in opposite directions, the atom interference loop formed by Raman pulse sequence is spatially symmetric \cite{NM2010}. By careful calculation we find that the initial phases of $k_{3}$ are canceled due to the opposite recoil process in the interference loop, and the phase shift of each site only depends on the initial phases $\varphi^{j}_{i0}$ of $k_{i} (i=1,2)$,\emph{ i.e.}
\begin{equation*}
\Delta\varphi_{j}=\varphi^{j}_{20}-\varphi^{j}_{10} (\emph{j}=A \sim D).
\end{equation*}
Similarly, for $^{87}$Rb atoms, the total phase shift of lasers is independent of $k_{4}$, and it is only sensitive to  $\varphi^{j}_{i0} (i=1,2)$. In other words, the 4WDR scheme is immune to phase noises of both $k_{3}$ and $k_{4}$. The residual noise of $\varphi^{j}_{i0} (i=1,2)$ can be common-mode rejected since $^{85}$Rb and $^{87}$Rb AIs share the same  $k_{1}$ and $k_{2}$.

The experimental setup \cite{Zhou2011} is a modified version of our early AIs \cite{Wang2007,LZhou2011}. Briefly, the magneto-optical trap (MOT) chamber is at bottom of the setup and on the top is the fountain pipe, in between is the detection chamber. A pair of rectangle windows for two parallel probe beams are arranged along the horizontal direction of the detection chamber. Two round windows (Window-A, Window-B) are perpendicular to the axial of the two rectangle windows, they are used for collecting laser-induced fluorescence from $^{85}$Rb and $^{87}$Rb simultaneously for each shot of fountain. Window-A is 30 mm higher than Window-B. All laser beams are supplied by the laser system, which is composed of a seed laser, a taped laser amplifier, and some acousto-optic modulators (AOMs). The seed laser is stabilized by saturated absorption spectroscopy and its frequency is shifted by AOMs. The blue detuning of Raman beams are realized by an electro-optic modulator \cite{Peng2014}.

Cold $^{85}$Rb and $^{87}$Rb atom clouds are prepared in MOT, and then launched simultaneously by a moving molasses process to form atom fountains. During the launching and falling process the 4WDR pulse sequence is applied. At the end $^{85}$Rb and $^{87}$Rb are detected parallelly at Window-A and Window-B. By scanning $\delta_{1}$ and $\delta_{2}$ simultaneously at chirp rates of $\alpha_{1}$ and $\alpha_{2}$ respectively the phase shifts of both $^{87}$Rb and $^{85}$Rb AIs are obtained. By switching the frequencies of two probe beams, $^{87}$Rb and $^{85}$Rb atoms in Windows-B and Windows-A are detected alternately.

\begin{figure}[htbp]
\centering
\includegraphics {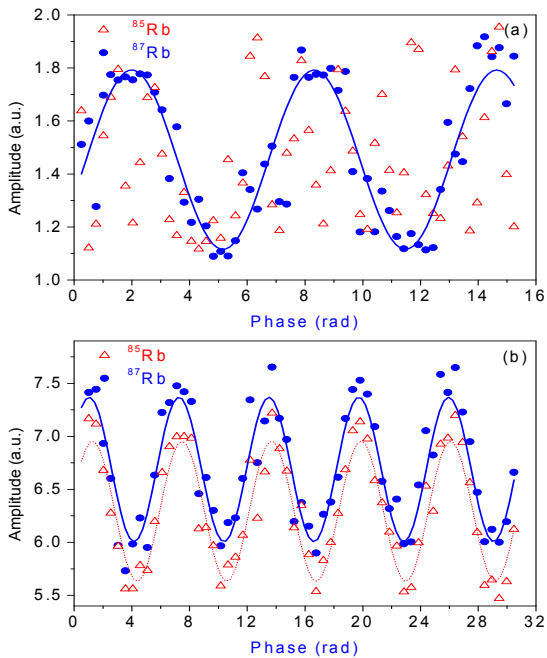}
\caption{(color online)Phase noise suppression by the 4WDR method. A rapid phase modulation is applied to $^{85}$Rb atoms. (a) Simultaneous $^{85}$Rb-$^{87}$Rb interference fringes obtained by single-diffraction Raman transition method and (b) Simultaneous $^{85}$Rb-$^{87}$Rb interference fringes by the 4WDR scheme. The red triangles are experimental data points of $^{85}$Rb atoms, and the red dotted line is sine curve fitting. The blue dots are experimental data points of $^{87}$Rb atoms, and the blue solid line is sine curve fitting. }
\label{fig:fig2}
\end{figure}

To evaluate the ability of phase noise suppression of the 4WDR scheme, a comparison experiment is performed. Firstly, we shut off the Raman beam with frequency of $\omega_{2}$, and carry out simultaneous $^{85}$Rb-$^{87}$Rb dual-species atom interferometry experiment by usual single-diffraction Raman transitions method. An AOM driven by a triangle wave is used to modulate the phase of $\omega_{3}$ to introduce rapid phase change to $^{85}$Rb atoms. The experimental data are shown in Fig.\ref{fig:fig2}(a). Due to the complicated phase variance from the modulation, $^{85}$Rb atom interference fringes disappears, while the visibility of unperturbed $^{87}$Rb atom interference fringes is 48\%. As a comparison, we then switch on the Raman beam of $\omega_{2}$, thus the AI is in a double-diffraction configuration. The visibility of $^{85}$Rb atom interference fringes as shown in Fig.\ref{fig:fig2}(b) is now about 20\% even it is still suffering from the phase modulation of $\omega_{3}$. This visibility is comparable with that of $^{87}$Rb atoms. Meanwhile as already demonstrated in \cite{TL2009,NM2010} the phase sensitivity of interference fringes obtained by the 4WDR method is improved by two times (see Fig.\ref{fig:fig2}(b)).

By using the 4WDR Raman AI we made gravity differential measurements. For each fringe we repeat 40 measurements, and a single measurement spends 2.5 s. By sine curve fitting we determine the chirp rates corresponding to the centers of fringes, they are $\alpha_{1}$ = 25.10408 MHz/s for $^{85}$Rb atoms and $\alpha_{2}$ = 25.10420 MHz/s for $^{87}$Rb atoms, respectively. The difference is mainly caused by the difference of effective wave vectors.

\begin{figure}[htbp]
\centering
\includegraphics{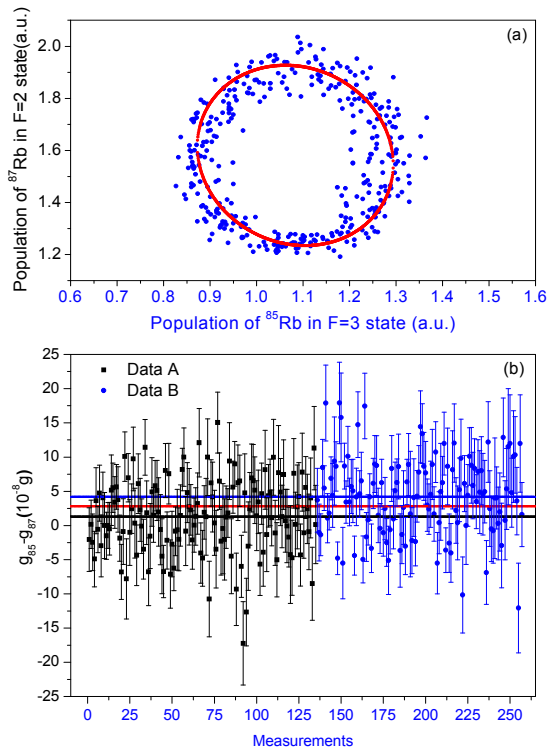}
\caption{(color online)Population of $^{87}$Rb in \emph{F}=2 state vs Population of $^{85}$Rb in F=3 state (a) and data for gravity differential measurements (b). The systematical error caused by difference of effective wave vectors of $^{85}$Rb and $^{87}$Rb is corrected. Data A are obtained by probing $^{87}$Rb atoms at Window-A while probing $^{85}$Rb atoms at Window-B; Data B are obtained by altering the probe position. The black line and the blue line are the average values of Data A and Data B respectively. The average of Data A and Data B is $2.8\times10^{-8}g$ shown as red line.}
\label{fig:fig3}
\end{figure}

To obtain phase difference between $^{85}$Rb and $^{87}$Rb simultaneous interference fringes, we conducted ellipse fitting by setting interference fringe data of $^{85}$Rb as \emph{x}, $^{87}$Rb as \emph{y} (see Fig.\ref{fig:fig3}(a) for a typical fringe data). For an ellipse fitting, the smallest error occurs if the data distribution is close to a perfect circle, where the phase difference is $\pi/2$. The value of $\Delta\varphi$ depends on experimental parameters \emph{T}, $k_{eff}$, $\alpha_{1}$, and $\alpha_{2}$. For given $k_{eff}$, $\alpha_{1}$, and $\alpha_{2}$, $\Delta\varphi$ can only be determined by \emph{T}. We set $T=70.96$ ms, and the corresponding fitted phase difference is near $\pi/2$. The frequency difference between $\omega_{3}$ and $\omega_{4}$ causes a systematic error of $-494.4\times10^{-8}g$ in gravity differential measurement.

We obtained two sets of data. Data A are fitted values of $g_{85}-g_{87}$ by probing $^{87}$Rb atoms at Window-A while probing $^{85}$Rb atoms at Window-B; Data B are $g_{85}-g_{87}$ by probing $^{85}$Rb at Window-A and $^{87}$Rb at Window-B. The average of Data A and Data B is $-491.6\times10^{-8}g$. Two sets of data after correcting system errors are shown in Fig.\ref{fig:fig3}(b), the average value of $g_{85}-g_{87}$ is $2.8\times10^{-8}g$. The relative gravity difference (namely, the E\"{o}tv\"{o}s parameter) can be obtained by
\begin{equation*}
\eta=\frac{(g_{85}-g_{87})}{(g_{85}+g_{87})/2}
\end{equation*}

\begin{figure}[htbp]
\centering
\includegraphics{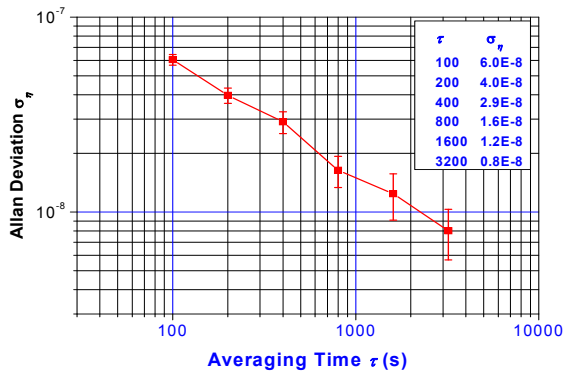}
\caption{(color online)Allan deviation of differential gravity measurement data of $^{85}$Rb and $^{87}$Rb atoms.The deviation of averaging 3200 s is $0.8\times10^{-8}$.}
\label{fig:fig4}
\end{figure}

Allan deviation of measurements for $\eta$ is shown in Fig.\ref{fig:fig4}. The deviation value $\sigma_{\eta}$ in the dual-logarithm chart decreases at the square root of averaging time $\tau$. At $\tau = 3200$ s the deviation is $0.8\times10^{-8}$. The well-behaved Allan deviation indicates that white noise is the dominant noise source in the experiment.
This again shows that the 4WDR scheme has good common-mode noise suppression ability at least as demonstrated here at the $10^{-8}$ level.

To give an uncertainty budget of errors other than the direct experimental measurement, \emph{i.e.} Type B errors, we make the following estimates.
The frequency difference between $\omega_{3}$ and $\omega_{4}$ is still a major systematic error, but because the uncertainty of laser frequency difference is less than 10 Hz, the uncertainty to correct the error is only $3\times10^{-11}$.
The fluctuation of bias magnetic field in our experiment is less than 1 mG, so the uncertainty of $\eta$ due to second order Zeeman shift is less than $1\times10^{-10}$.
Due to the tiny but not zero difference of $^{85}$Rb and $^{87}$Rb atoms in mass, launch velocity and recoil velocity, the central positions of two species atom clouds are not completely overlapped during the free falling process. The Coriolis effect caused by Earth's rotation coupling with free falling atoms due to their horizontal velocity distribution, the fluctuations of initial positions and velocities of two species atoms, is another uncertainty source of the E\"{o}tv\"{o}s parameter. The uncertainty of the horizontal position difference of two clouds is less than 2 mm, and the uncertainty of velocity difference is less than 1 mm/s. Considering the latitude of our laboratory (north latitude $30.54^{\circ}$), the calculated uncertainty caused by Coriolis effect is $2.9\times10^{-8}$.
The vertical position difference of $^{85}$Rb and $^{87}$Rb atom clouds is $0.23\pm1.00$ mm, thus the gravity gradient based systematic error is less than $7\times10^{-11}$, and its uncertainty is $3\times10^{-10}$.
In our experiments, the fluctuation of laser intensities is less than 10\%, the uncertainty of $\eta$ due to AC Stark shifts is measured in independent experiments to be less than $2\times10^{-9}$.

\begin{table}[htbp]
\caption{\label{tab:table1}%
 Main contributions affecting the differential gravitational acceleration measurement.
}
\begin{ruledtabular}
\begin{tabular}{lcdr}
\multicolumn{1}{c}{\textrm{}}&
\textrm{$\eta(\times10^{-8})$}&
\textrm{Uncertainty($\times10^{-8}$)}&\\
\colrule
\multicolumn{1}{c}{\textrm{Experimental data}}& \textrm{-491.6}& \textrm{0.8}&\\
\colrule
\multicolumn{1}{c}{\textrm{Effective wave vector}}& \textrm{-494.4}& \textrm{0.0}&\\
\multicolumn{1}{c}{\textrm{Second order Zeeman shift}}& \textrm{0}& \textrm{0.01}&\\
\multicolumn{1}{c}{\textrm{Gravity gradient}}& \textrm{0.01}&\textrm{0.03}&\\
\multicolumn{1}{c}{\textrm{Coriolis effect}}& \textrm{0}& \textrm{2.9}&\\
\multicolumn{1}{c}{\textrm{AC Stark shift}}& \textrm{0}& \textrm{0.2}&\\
\colrule
\multicolumn{1}{c}{\textrm{Total}}& \textrm{2.8}& \textrm{3.0}&\\
\end{tabular}
\end{ruledtabular}
\end{table}

All above mentioned main contributions affecting the differential acceleration measurement are listed in Table \ref{tab:table1}. Including all statistical uncertainties or errors (Type A and B) together, the total uncertainty of $\eta$ value is $3.0\times10^{-8}$.

To further reduce the uncertainty, Coriolis effect should be canceled. It can be done by rotating the mirrors \cite{SYL2012} reflecting Raman beams. Then the signal to noise ratio should be increased in our experiment by evolving more and further cooled atoms, and by suppressing residual noises like seismic vibration with active vibration-isolation \cite{Tang2014}. Finally the 10-meter fountain AIs \cite{Zhou2011,SMD2013} or even AI in space \cite{DNA2014} will come to play with their ultrahigh sensitivity.

In summary, we developed a simultaneous dual-species ($^{85}$Rb-$^{87}$Rb) cold AI in which the proposed 4WDR scheme was used and demonstrated to have obvious advantages in immunizing common-mode noises. The 4WDR AI carries forward all features, including larger interference loop, better phase sensitivity and suppression of the phase noises of external fields, revealed in single species counterpart. It also holds the new ability of suppressing common-mode phase noise of Raman lasers in dual-species case. With this new type AI, we made a new WEP test at $10^{-8}$ level and found no violation to the WEP. This work advances a step forward in WEP test with atoms by improving the accuracy about one order.

We acknowledge Jun Luo, Tianchu Li and Jun Ye for their helpful discussions and suggestions in error evaluation and data analysis. This work was supported by the National Basic Research Program of China under Grant No. 2010CB832805, by the National Natural Science Foundation of China under Grant Nos.11227803 and 91436107, and also by funds from the Chinese Academy of Sciences.

\end{document}